\documentclass[aps,pra,reprint,showpacs,superscriptaddress,notitlepage,twocolumn]{revtex4-1}%
\usepackage{amsfonts}
\usepackage{mathrsfs}
\usepackage{amsmath}
\usepackage{amssymb}
\usepackage{graphicx}
\usepackage{color}%
\usepackage{mathtools}
\usepackage{booktabs}
\usepackage[colorlinks,linkcolor=blue,citecolor=blue,hyperindex,bookmarks=false,pdfstartview=FitH]{hyperref}

\providecommand{\U}[1]{\protect\rule{.1in}{.1in}}

\newcommand{\figpanel}[2]{\hyperref[#1]{\ref*{#1}(#2)}}

\begin{document}
\title{Giant atoms with time-dependent couplings}

\author{Lei Du}
\affiliation{Beijing Computational Science Research Center, Beijing 100193, China}
\author{Yao-Tong Chen}
\affiliation{Center for Quantum Sciences and School of Physics, Northeast Normal University, Changchun 130024, China}
\author{Yan Zhang}
\email{zhangy345@nenu.edu.cn}
\affiliation{Center for Quantum Sciences and School of Physics, Northeast Normal University, Changchun 130024, China}
\author{Yong Li}
\email{liyong@csrc.ac.cn}
\affiliation{Beijing Computational Science Research Center, Beijing 100193, China}
\affiliation{Center for Theoretical Physics and School of Science, Hainan University, Haikou 570228, China}
\affiliation{Synergetic Innovation Center for Quantum Effects and Applications, Hunan Normal University, Changsha 410081, China}

\date{\today }

\begin{abstract}
We study the decay dynamics of a two-level giant atom that is coupled to a waveguide with time-dependent coupling strengths. In the non-Markovian regime where the retardation effect cannot be ignored, we show that the dynamics of the atom depends on the atom-waveguide coupling strengths at an earlier time. This allows one to tailor the decay dynamics of the giant atom and even realize a stationary population revival with appropriate coupling modulations. Moreover, we demonstrate the possibility of simulating the quantum Zeno and quantum anti-Zeno effects in the giant-atom model with periodic coupling quenches. These results have potential applications in quantum information processing and quantum network engineering.    
\end{abstract}

\maketitle


\section{Introduction}

Giant atoms have spurred a rapidly growing interest in the past few years due to the exotic self-interference effects therein~\cite{AFKbook}. Such systems feature nonlocal interactions between the atoms and the waveguide fields, which are possible if the atomic size is much larger than the wavelength of the field~\cite{SAW1,SAW2,SAW3,SAW4,SAW5,SAW6} or if the field is confined in a meandering waveguide that can contact with each atom multiple times~\cite{Lamb,braided,EngineerCW}. While the interaction at each atom-field coupling point can still be well described by the dipole approximation, the atoms in these systems can no longer be viewed as points and the phase accumulations of photons (or phonons) between different coupling points should be taken into account. To date, there have been a variety of intriguing phenomena witnessed in giant-atom structures, such as frequency-dependent Lamb shifts and relaxation rates~\cite{Lamb}, decoherence-free interatomic interactions~\cite{braided,GANori,AFKchiral,DFmechanism,CollisionFC}, unconventional bound states~\cite{WXbound,oscillate,osci2,ZhaoGA,ChengTopo,TudelaGA,YuanGA}, and phase-controlled frequency conversions~\cite{DLlambda,DLprr}. Most recently, giant atoms have also been extended to the non-perturbative regime~\cite{Nperturbative}, to chiral quantum optics~\cite{DFmechanism,CollisionFC,WXbound,Karg2019,AFKchiral,WXchiral}, and to synthetic dimensions~\cite{DLsyn}.  

Besides the progress above, it is also an interesting topic to study non-Markovian retardation effects in giant-atom systems. Indeed, such effects are common, and should be taken into account if the propagation time of photons between different coupling points is comparable to or even larger than the lifetime of the atom~\cite{SAW5}. In this case, both the dynamic evolutions~\cite{SAW5,GLZ2017,oscillate,DLretard} and the stationary scattering properties~\cite{DLprr,ZhuGA,JiaGA} of the giant atom exhibit significant non-Markovian features that have no counterparts in the Markovian regime, such as bound states that oscillate persistently between the giant atom and the one-dimensional continuum~\cite{oscillate} and non-Markovianity induced nonreciprocity~\cite{DLprr}. Moreover, non-Markovian retardation effects have also been well studied in systems featuring a semi-infinite waveguide, where a small atom placed in front of the waveguide end can be mapped into a giant atom with two identical coupling strengths~\cite{SemiDH,SemiShen,FCretard1,FCretard2,FCretard3}. In these works, however, the atom-field couplings are assumed to be constant and the non-Markovian retardation effects affect the dynamics in a relatively simple manner.      

In this paper, we consider a two-level giant atom that is coupled to a waveguide with time-dependent coupling strengths. We reveal that the non-Markovian retardation effect is closely related to the instantaneous atom-waveguide coupling strengths at an earlier time. Based on this mechanism, we consider some simple modulation schemes for the time-dependent couplings, which enable dynamical control of the decay dynamics of the giant atom without changing the frequency of the emitted photons. In particular, it is possible to observe a limited energy backflow from the waveguide to the atom via a sudden change in the coupling strengths. Moreover, we demonstrate how to simulate the quantum Zeno effect (QZE) and quantum anti-Zeno effect (QAZE)~\cite{Zeno1977,AZEnature,FacchiPRL2001,FacchiPRA2004,FacchiJPA} through a sequence of coupling quenches. This provides an alternative platform for studying Zeno physics and quench dynamics in open quantum systems.  

\section{Model and equations}

\begin{figure}[ptb]
\centering
\includegraphics[width=8.5 cm]{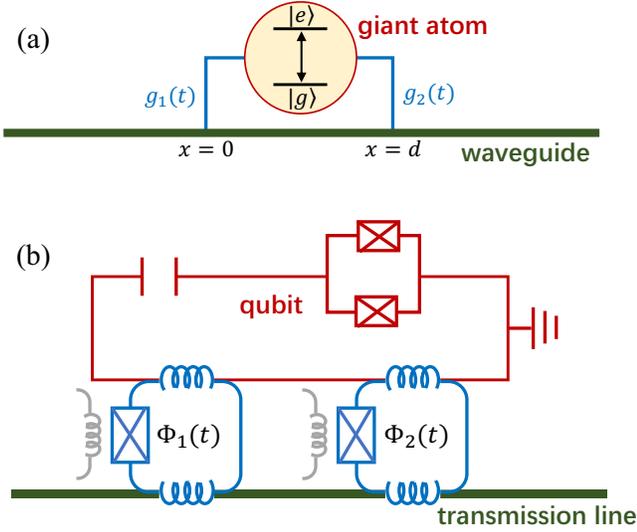}
\caption{(a) Schematic illustration of a two-level giant atom coupled to a waveguide with time-dependent coupling strengths $g_{1}(t)$ and $g_{2}(t)$. (b) A physical implementation of the model in (a) with superconducting quantum elements, where a transmon is coupled to a transmission line via two Josephson loops. The time-dependent couplings can be achieved by modulating the external fluxes $\Phi_{1}(t)$ and $\Phi_{2}(t)$ in time.}\label{fig1}
\end{figure}

We consider a two-level giant atom, which is coupled to a waveguide at two different points with time-dependent coupling strengths $g_{1}(t)$ and $g_{2}(t)$, respectively, as shown in Fig.~\figpanel{fig1}{a}. For ultracold atoms in optical lattices, time-dependent couplings can be implemented by dynamically modulating the relative position of the potentials. Such a scheme has recently been used to simulate an effective giant atom coupled to a high-dimensional bath~\cite{AGTGA}. Here we consider a viable solid-state implementation platform based on superconducting quantum circuits. As shown in Fig.~\figpanel{fig1}{b}, a transmon qubit is coupled to a superconducting transmission line with the interaction mediated by a Josephson loop (i.e., a loop containing a Josephson junction) at each coupling point~\cite{WXchiral}. In this way, the time-dependent couplings can be achieved by modulating the external fluxes through the loops~\cite{modu1,modu2,modu3,modu4}. In this case, the Hamiltonian of the system can be written as ($\hbar=1$ hereafter)
\begin{equation}
\begin{split}
H&=\omega_{0}\sigma_{+}\sigma_{-}+\int_{-\infty}^{+\infty}dk \omega_{k}a_{k}^{\dag}a_{k}\\
&\quad\,+\int_{-\infty}^{+\infty}dk\Big\{\Big[g_{1}(t)+g_{2}(t)e^{ikd}\Big]\sigma_{+}a_{k}+\text{H.c.}\Big\},
\end{split}
\label{eq1}
\end{equation} 
where $\sigma_{+}$ ($\sigma_{-}$) is the raising (lowering) operator of the giant atom with transition frequency $\omega_{0}$; $a_{k}^{\dag}$ ($a_{k}$) is the creation (annihilation) operator of the waveguide mode with frequency $\omega_{k}$ and wave vector $k$ (along the propagation direction); $d$ is the separation distance between the two atom-waveguide coupling points. In Eq.~(\ref{eq1}), we have employed the Weisskopf-Wigner approximation~\cite{walls} (the intensity of the atomic power spectrum is concentrated at the transition frequency $\omega_{0}$) such that the coupling strengths $g_{1,2}(t)$ can be treated as $k$ independent. In the single-excitation subspace, the state of the system at time $t$ has the form
\begin{equation}
|\psi(t)\rangle=\Big[\int_{-\infty}^{+\infty}dkc_{k}(t)a_{k}^{\dag}e^{-i\omega_{k}t}+c_{e}(t)\sigma_{+}e^{-i\omega_{0}t}\Big]|V\rangle,
\label{eq2}
\end{equation} 
where $c_{k}(t)$ [$c_{e}(t)$] is the probability amplitude of creating a photon with wave vector $k$ in the waveguide (of exciting the giant atom); $|V\rangle$ denotes the ground state of the whole system. By solving the Schr\"{o}dinger equation, one obtains 
\begin{equation}
\begin{split}
\dot{c}_{e}(t)&=-i\int_{-\infty}^{+\infty}dk\Big[g_{1}(t)+g_{2}(t)e^{ikd}\Big]\\
&\quad\,\times c_{k}(t)e^{-i(\omega_{k}-\omega_{0})t},\\
\dot{c}_{k}(t)&=-i\Big[g_{1}(t)+g_{2}(t)e^{-ikd}\Big]c_{e}(t)e^{i(\omega_{k}-\omega_{0})t},
\end{split}
\label{eq3}
\end{equation}
where the superscript dot denotes the derivative with respect to time $t$. For the case that the waveguide field is initialized in the vacuum state, the formal solution of $c_{k}(t)$ can be written as
\begin{equation}
\begin{split}
c_{k}(t)&=-i\int_{0}^{t}dt'[g_{1}(t')+g_{2}(t')e^{-ikd}]\\
&\quad\,\times c_{e}(t')e^{i(\omega_{k}-\omega_{0})t'}.
\end{split}
\label{eq4}
\end{equation}
Substituting Eq.~(\ref{eq4}) into the dynamic equation of $c_{e}(t)$ in Eq.~(\ref{eq3}), one has
\begin{equation}
\begin{split}
\dot{c}_{e}(t)&=-2\int_{0}^{t}dt'\int_{0}^{\infty}\Big\{g_{1}(t)g_{1}(t')+g_{2}(t)g_{2}(t')\\
&\quad\,\,+[g_{1}(t)g_{2}(t')+g_{1}(t')g_{2}(t)]\cos{(kd)}\Big\}\\
&\quad\,\,\times c_{e}(t')e^{-i(\omega_{k}-\omega_{0})(t-t')}\frac{d\omega_{k}}{v_{g}},
\end{split}
\label{eq5}
\end{equation}
where we have changed the integration variable as $\int_{-\infty}^{+\infty}dk=2\int_{0}^{+\infty}d\omega_{k}/v_{g}$ with $v_{g}$ the group velocity of the emitted photons in the waveguide. According to the Weisskopf-Wigner approximation, one can assume $\omega_{k}\approx\omega_{0}+\nu=\omega_{0}+(k-k_{0})v_{g}$ in the vicinity of $\omega_{0}$, with $k_{0}$ the wave vector corresponding to $\omega_{0}$~\cite{shen2005,shen2009}. In this way, Eq.~(\ref{eq5}) becomes (see Appendix~\ref{appa} for more details) 
\begin{equation}
\begin{split}
\dot{c}_{e}(t)
&=-\frac{1}{2}[\Gamma_{1}(t,0)+\Gamma_{2}(t,0)]c_{e}(t)-\Gamma_{12}(t,\tau)e^{i\phi}\\
&\quad\,\times c_{e}(t-\tau)\Theta(t-\tau),
\end{split}
\label{eq6}
\end{equation}
where $\phi=k_{0}d$ and $\tau=d/v_{g}$ are the phase accumulation and the propagation time of photons traveling between the two atom-waveguide coupling points, respectively; $\Gamma_{j}(t,0)=4\pi g_{j}(t)^{2}/v_{g}$ ($j=1,2$) is the instantaneous decay rate of the atom at the $j$th coupling point and $\Gamma_{12}(t,\tau)=2\pi [g_{1}(t)g_{2}(t-\tau)+g_{1}(t-\tau)g_{2}(t)]/v_{g}$ describes the retarded correlation decay due to the giant-atom structure; $\Theta(t)$ is the Heaviside step function. 

Equation~(\ref{eq6}) shows that in the non-Markovian regime, where the propagation time $\tau$ is comparable to or larger than the lifetime of the atom, the retarded feedback term (i.e., the second term) depends on the coupling strengths $g_{1}(t-\tau)$ and $g_{2}(t-\tau)$ of the earlier moment $t-\tau$. We point out that such a non-Markovian feature is crucial for observing the unconventional results demonstrated below. For the case of time-independent coupling strengths $g_{1}(t)\equiv g_{1}$ and $g_{2}(t)\equiv g_{2}$, Eq.~(\ref{eq6}) reduces to
\begin{equation}
\dot{c}_{e}(t)=-\frac{1}{2}(\Gamma_{1}+\Gamma_{2})c_{e}(t)-\Gamma_{12}e^{i\phi}c_{e}(t-\tau)\Theta(t-\tau)
\label{eq7}
\end{equation}
with $\Gamma_{j}=4\pi g_{j}^{2}/v_{g}$ and $\Gamma_{12}=4\pi g_{1}g_{2}/v_{g}$. If $g_{1}=g_{2}$, Eq.~(\ref{eq7}) is strictly connected to the dynamic equation governing a small atom coupled in a time-independent manner to a semi-infinite waveguide~\cite{FCretard1,FCretard2,FCretard3}. In the following, we will investigate how the time dependence of the coupling strength affects the decay dynamics of the giant atom.

\section{Cosine-shaped modulation}

We first consider cosine-shaped modulations for the atom-waveguide coupling strengths, i.e., $g_{j}(t)=g_{j,0}\cos{(\Omega_{j}t+\theta_{j})}$, with $g_{j,0}$, $\Omega_{j}$, and $\theta_{j}$ the amplitude, frequency, and initial phase of the modulation at the $j$th coupling point, respectively. Here we restrict ourselves to the simple case of $g_{1,0}=g_{2,0}=g_{0}$ and $\Omega_{1}=\Omega_{2}=\Omega$. We only consider an initial phase difference by assuming $\theta_{1}=0$ and $\theta_{2}=\theta$ without loss of generality. In this case Eq.~(\ref{eq6}) becomes
\begin{equation}
\begin{split}
\dot{c}_{e}(t)&=-\frac{\Gamma_{0}}{2}[\cos^{2}{(\Omega t)}+\cos^{2}{(\Omega t+\theta)}]c_{e}(t)\\
&\quad\,-\frac{\Gamma_{0}}{2}e^{i\phi}[\cos{(\Omega t)}\cos{(\Omega t-\Omega\tau+\theta)}\\
&\quad\,+\cos{(\Omega t-\Omega\tau)}\cos{(\Omega t+\theta)}]c_{e}(t-\tau)\Theta(t-\tau)
\end{split}
\label{eq8}
\end{equation}    
with $\Gamma_{0}=4\pi g_{0}^{2}/v_{g}$. Clearly, the dynamics of the giant atom can be controlled via either the modulation frequency $\Omega$ or the initial phase difference $\theta$, both of which are experimentally tunable. For example, Eq.~(\ref{eq8}) can be simplified to
\begin{equation}
\begin{split}
\dot{c}_{e}(t)&=-\frac{\Gamma_{0}}{2}\Big\{[\cos^{2}{(\Omega t)}+\cos^{2}{(\Omega t+\theta)}]c_{e}(t)\\
&\quad\,-2\cos{(\Omega t)}\cos{(\Omega t+\theta)}e^{i\phi}c_{e}(t-\tau)\Theta(t-\tau)\Big\}
\end{split}
\label{eq9}
\end{equation}
if $\Omega\tau=(2m+1)\pi$ ($m$ is an arbitrary integer), to
\begin{equation}
\begin{split}
\dot{c}_{e}(t)&=-\frac{\Gamma_{0}}{2}\Big\{[\cos^{2}{(\Omega t)}+\cos^{2}{(\Omega t+\theta)}]c_{e}(t)\\
&\quad\,+2\cos{(\Omega t)}\cos{(\Omega t+\theta)}e^{i\phi}c_{e}(t-\tau)\Theta(t-\tau)\Big\}
\end{split}
\label{eq10}
\end{equation}
if $\Omega\tau=2m\pi$, and to
\begin{equation}
\begin{split}
\dot{c}_{e}(t)&=-\frac{\Gamma_{0}}{2}\Big\{[\cos^{2}{(\Omega t)}+\cos^{2}{(\Omega t+\theta)}]c_{e}(t)\\
&\quad\,+e^{i\phi}[\cos{(\Omega t)}\sin{(\Omega t+\theta)}\\
&\quad\,+\sin{(\Omega t)}\cos{(\Omega t+\theta)}]c_{e}(t-\tau)\Theta(t-\tau)\Big\}
\end{split}
\label{eq11}
\end{equation}
if $\Omega\tau=(2m+1/2)\pi$. In this case, further control of the atomic dynamics can be achieved by tuning the initial phase difference $\theta$ as mentioned above.      

\begin{figure}[ptb]
\centering
\includegraphics[width=8.5 cm]{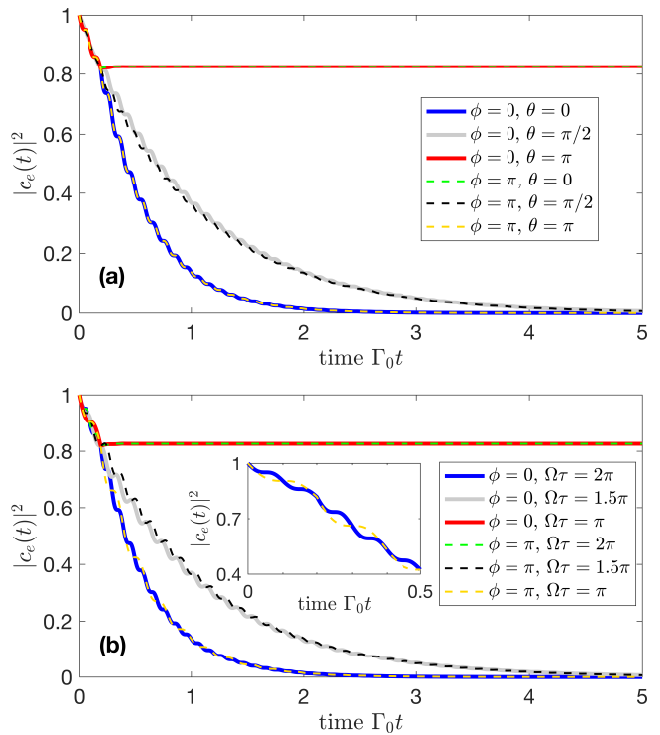}
\caption{Dynamic evolutions of atomic population $|c_{e}(t)|^{2}$ with initial state $|\psi(0)\rangle=\sigma_{+}|V\rangle$ and cosine-shaped modulations. We assume $\Omega/\Gamma_{0}=10\pi$ in (a) and $\theta=0$ in (b). The inset in (b) is the enlarged view of chosen lines. The other parameters are $\Gamma_{0}=4\pi g_{0}^{2}/v_{g}$ and $\Gamma_{0}\tau=0.2$.}\label{fig2}
\end{figure}

Before proceeding, we briefly revisit the typical dynamics of a time-independent giant atom described by Eq.~(\ref{eq7}). It has been shown that the spontaneous emission of such an atom can be suppressed if $\phi=(2m+1)\pi$ and $g_{1}=g_{2}$ (the two atom-waveguide coupling paths interfere destructively) and if the propagation time $\tau$ is negligible compared with the lifetime of the atom (i.e., the system is in the Markovian regime)~\cite{Lamb,GLZ2017,SAW5,FCretard1,FCretard2,FCretard3}. In this case, the giant atom is effectively decoupled from the waveguide and becomes ``decoherence free''~\cite{braided,GANori}. On the other hand, the atom can also exhibit superradiance behavior if the two coupling paths interfere constructively (in this context, ``superradiance'' refers to enhanced radiative decay due to the giant-atom structure~\cite{LonghiGA}). 

For the time-dependent model here, we demonstrate in Fig.~\ref{fig2} the dynamic evolutions of the atomic population $|c_{e}(t)|^{2}$ (the atom is in the excited state initially) with different values of $\Omega\tau$ and $\theta$. If both $\Omega\tau$ and $\theta$ are integer multiples of $2\pi$, as shown in Fig.~\figpanel{fig2}{a}, the atom still exhibits the long-lived population and the superradiance emission when $\phi=(2m+1)\pi$ and $\phi=2m\pi$, respectively, similar to the case with time-independent coupling strengths. In this case, the superradiance behavior shows a slight oscillation arising from the cosine-shaped modulations, while the long-lived population still does not change with time since the two decay rates (i.e., the instantaneous and the retarded decay rates) are always identical. Interestingly, it shows that the decay dynamics of the atom can be tuned flexibly between the long-lived population and the superradiance behavior by changing the phase difference $\theta$. This can be understood from Eq.~(\ref{eq8}): the retarded feedback term is completely opposite when $\theta=0$ and $\theta=\pi$, while its influence on the decay dynamics is halfway between the two extremes when $0<\theta<\pi$. 
In view of this, the present proposal provides an alternative scheme for tailoring the decay dynamics of the atom without changing its transition frequency~\cite{SAW5}. This scheme is, however, not suitable for a small atom in front of a mirror because it is challenging to introduce the phase difference $\theta$ in this case.      

The product $\Omega\tau$ of the modulation frequency and the propagation time also plays an important role for the decay dynamics, as shown in Eq.~(\ref{eq8}). This can be verified from the results in Fig.~\figpanel{fig2}{b}, where we change the value of $\Omega$ and fix the propagation time $\tau$ (i.e., the separation between the coupling points). In this case, the decay dynamics can be tuned between the decoherence-free behavior and the superradiance behavior by changing the value of $\Omega$ instead. This can also be seen from the opposite signs of the feedback terms in Eqs.~(\ref{eq9}) and (\ref{eq10}). However, the superradiance behaviors are no longer coincident when $\phi$ takes different values (see the blue solid and yellow dashed lines). The latter one shows a slower oscillation due to the smaller modulation frequency, as shown in the inset in Fig.~\figpanel{fig2}{b}.   

\section{Steplike modulation}

\begin{figure}[ptb]
\centering
\includegraphics[width=8.5 cm]{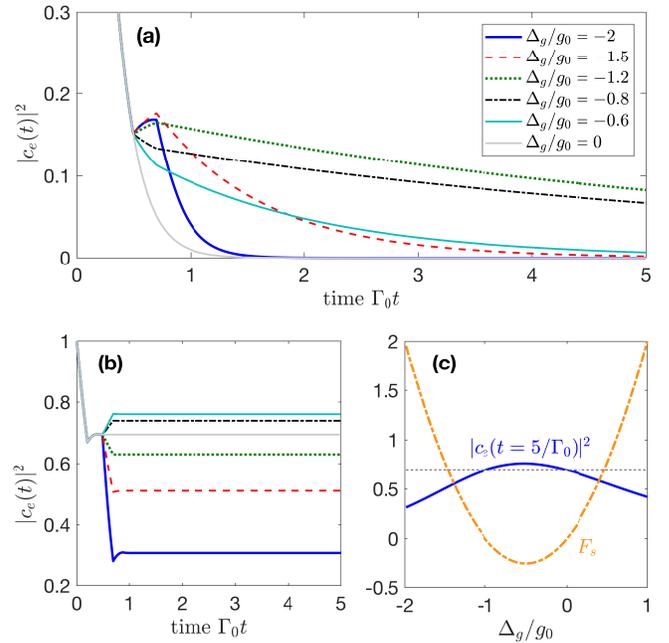}
\caption{(a) and (b) Dynamic evolutions of atomic population $|c_{e}(t)|^{2}$ with initial state $|\psi(0)\rangle=\sigma_{+}|V\rangle$ and steplike modulations. We assume $\phi=0$ in (a) and $\phi=\pi$ in (b). (a) and (b) share the same legend for the values of $\Delta_{g}/g_{0}$. (c) Atomic population $|c_{e}(t)|^{2}$ at time $t=5/\Gamma_{0}$ and the value of $F_{s}$ versus $\Delta_{g}/g_{0}$ when $\phi=\pi$. The horizontal gray dashed line in (c) corresponds to the value of $|c_{e}(t=5/\Gamma_{0})|^{2}$ without modulation ($\Delta_{g}/g_{0}=0$). Other parameters are $\Gamma_{0}=4\pi g_{0}^{2}/v_{g}$, $\Gamma_{0}\tau=0.2$, and $\Gamma_{0}t'=0.5$.}\label{fig3}
\end{figure}

In this section, we would like to consider steplike modulations for the atom-waveguide coupling strengths, i.e., the coupling strengths change abruptly at a specific moment from one value to another. For simplicity, we assume identical modulations at the two coupling points, i.e., $g_{1}(t)=g_{2}(t)=g_{0}+\Delta_{g}\Theta(t-t')$ in Eq.~(\ref{eq6}) with $g_{0}$ and $\Delta_{g}$ the initial value and the variation of the coupling strengths, respectively. 

Figures~\figpanel{fig3}{a} and \figpanel{fig3}{b} depict the dynamic evolutions of the atomic population with the steplike modulations for $\phi=0$ and $\phi=\pi$, respectively. In both cases, one can find a revival or reduction of the population from $t=t'$ to $t=t'+\tau$. For $\phi=2m\pi$, as shown in Fig.~\figpanel{fig3}{a}, the atomic population regains damped when $t>t'+\tau$, with the damping rate determined by the absolute value of the final coupling strengths. For $\phi=(2m+1)\pi$, however, the atomic population becomes undamped again after the variation, as shown in Fig.~\figpanel{fig3}{b}. This implies that one can partially offset the initial energy loss of the giant atom that arises from the non-Markovian retardation effect.

The results in Fig.~\figpanel{fig3}{b} can be understood from Eq.~(\ref{eq6}) for $t\in[t',t'+\tau)$, i.e.,
\begin{equation}
\dot{c}_{e}(t)=-\frac{4\pi(g_{0}+\Delta_{g})^{2}}{v_{g}}c_{e}(t)-\frac{4\pi g_{0}(g_{0}+\Delta_{g})}{v_{g}}e^{i\phi}c_{e}(t-\tau),
\label{eq12}
\end{equation} 
and that for $t>t'+\tau$, i.e.,
\begin{equation}
\dot{c}_{e}(t)=-\frac{4\pi(g_{0}+\Delta_{g})^{2}}{v_{g}}\Big[c_{e}(t)+e^{i\phi}c_{e}(t-\tau)\Big].
\label{eq13}
\end{equation} 
On one hand, Eq.~(\ref{eq12}) shows that for $\phi=(2m+1)\pi$ a population revival (reduction) should occur at $t=t'$ if the value of 
\begin{equation}
F_{s}\coloneqq\frac{\Delta_{g}}{g_{0}}\Big(1+\frac{\Delta_{g}}{g_{0}}\Big)
\label{eq14}
\end{equation}
is negative (positive) (see Appendix~\ref{appb} for more details). This can be seen in Fig.~\figpanel{fig3}{c}, where the steady value of the atomic population is inversely proportional to $F_{s}$, and if $F_{s}<0$, it becomes larger than that without modulation (see the horizontal dashed line). On the other hand, for $t>t'+\tau$, one can see from Eq.~(\ref{eq13}) that the two decay paths cancel each other again if $\phi=(2m+1)\pi$. The results in Fig.~\figpanel{fig3}{a} are much more complicated to analyze with this approach. Nevertheless, the sudden revival or reduction and the modified decay of the atom can also be explained by the altered interference effect between the instantaneous and retarded terms in Eq.~(\ref{eq6}). 

We would like to point out that the population revival in Fig.~\ref{fig3} does not violate the energy conservation of the whole system. According to Eq.~(\ref{eq12}), the sudden change in the coupling strength $g(t)$ modifies the interference effect of the instantaneous and retarded terms, which leads to a further energy loss from the atom to the waveguide, or an energy backflow from the waveguide (the region between the two coupling points) to the atom. However, the atomic population cannot grow back to unity due to the initial energy loss before $t=\tau$.  
 
\section{Quantum Zeno and quantum anti-Zeno effects}       
 
Considering that the subsequent decay of the atom can be tuned by a sudden change in the atom-waveguide coupling strengths, as shown in Fig.~\ref{fig3}, it is natural to ask if the QZE and its inverse version, i.e., QAZE, can be simulated by repeating such sudden changes. The answer is positive, as will be shown below. The QZE (also known as the Zeno's paradox) states that the decay of an unstable quantum system can be hindered by frequent observations. This effect requires that the time interval of the observations is shorter than the Zeno time before which the survival probability of the system exhibits a short-time quadratic decay~\cite{Zeno1977,AZEnature,FacchiPRL2001,FacchiPRA2004,FacchiJPA}. Such a short-time behavior, however, is lacking in dynamics governed by Eq.~(\ref{eq6}) since the Weisskopf-Wigner approximation kills the short-time memory of the environment (the waveguide). In view of this, we resort to a discrete version of the present model, where a two-level giant atom is coupled to a tight-binding lattice (e.g., a one-dimensional array of coupled transmission line resonators~\cite{TLR1,TLR2}) with time-dependent coupling strengths. In this case, the short-time behavior can be clearly observed by directly solving the coupled-mode equations of the whole system~\cite{LonghiZeno2006,OhZeno}. Now the Hamiltonian of the system can be written as
\begin{equation}
\begin{split}
H'&=\omega_{0}\sigma_{+}\sigma_{-}+\sum_{m}\Big[\omega_{a}a_{m}^{\dag}a_{m}-J(a_{m}^{\dag}a_{m+1}\\
&\quad\,+\text{H.c.})\Big]+\Big[g(t)\sigma_{+}(a_{0}+a_{N})+\text{H.c.}\Big],
\end{split}
\label{eq15}
\end{equation}
where $a_{m}$ is the annihilation operator of the $m$th resonator of the lattice; $\omega_{a}$ is the frequency of each resonator in the array; $J$ is the coupling constant between adjacent resonators (for simplicity, we only consider the nearest-neighbor couplings). Here we have assumed that the atom is coupled to the $0$th and the $N$th resonators of the lattice with identical time-dependent coupling strength $g(t)$. By performing the transformation $a_{m}=\sum_{k}a_{k}e^{ikm}/\sqrt{2\pi}$, Eq.~(\ref{eq15}) becomes
\begin{equation}
\begin{split}
H'&=\omega_{0}\sigma_{+}\sigma_{-}+\int dk\tilde{\omega}_{k}a_{k}^{\dag}a_{k}\\
&\quad\,+\int dk\Big[\tilde{g}(t)(1+e^{ikN})\sigma_{+}a_{k}+\text{H.c.}\Big]
\end{split}
\label{eq16}
\end{equation}
with $\tilde{\omega}_{k}=\omega_{a}-2J\cos{k}$ the dispersion relation of the lattice and $\tilde{g}(t)=g(t)/\sqrt{2\pi}$ the renormalized coupling strength. In view of this, the tight-binding model here serves as a one-dimensional structured waveguide: If the atomic frequency is within the energy band of the lattice, i.e., $\omega_{0}\in[\omega_{a}-2J,\omega_{a}+2J]$, Eq.~(\ref{eq16}) basically describes a giant atom weakly coupled to a bath where photon escape from the atom to the lattice can be observed; otherwise, if the atom is tuned off-resonance to the lattice band, the photon escape is prohibited and atom-photon bound states can be formed~\cite{PAbound1,PAbound2,PAbound3}. In the resonant case of $\omega_{0}=\omega_{a}$ (i.e., the frequency of the atom lies at the middle of the energy band of the lattice), the phase accumulation between the two atom-lattice coupling points is given by $\phi=N\pi/2$~\cite{LonghiGA,Longhiparadox}. In this case, the dynamic evolution of the atomic population $|c_{e}(t)|^{2}$ can be determined by solving the coupled-mode equations
\begin{equation}
\begin{split}
\dot{c_{e}}&=-ig(t)(c_{0}+c_{N}),\\
\dot{c_{m}}&=-iJ(c_{m-1}+c_{m+1})-ig(t)c_{e}(\delta_{m,0}+\delta_{m,N}).
\end{split}
\label{eq17}
\end{equation}
For our purpose here, we consider a periodic quench scheme 
\begin{equation}
\begin{split}
g(t)&=g_{0}\Big\{\sum_{n}\Theta[t-(n-1)t'-(n-1)t'']\\
&\quad\,-\sum_{n}\Theta[t-nt'-(n-1)t'']\Big\}
\end{split}
\label{eq18}
\end{equation} 
for the time-dependent coupling strength in Eq.~(\ref{eq15}) with $n\in\mathbb{Z}^{+}$, where $t'$ and $t''$ are the turn-on and quench-off durations within each period, respectively.

\begin{figure}
\includegraphics[width=8.5 cm]{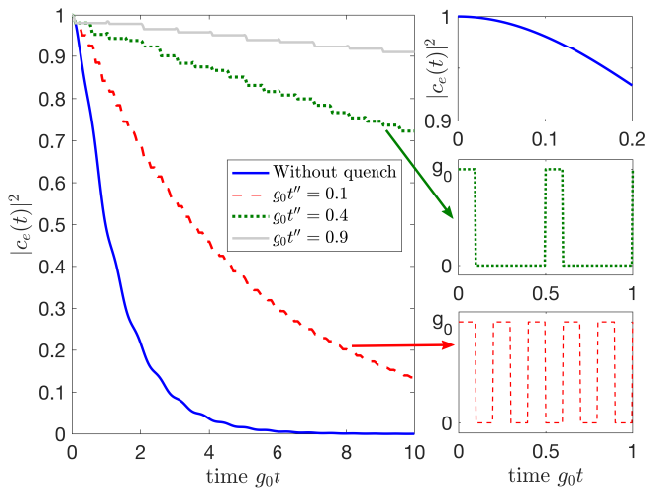}
\caption{Dynamic evolutions of atomic population $|c_{e}(t)|^{2}$ with initial state $|\psi(0)\rangle=\sigma_{+}|V\rangle$ in the case of a discrete waveguide. The insets depict, from top to bottom, the enlarged view of the chosen (blue solid) line in the main figure, the evolutions of the time-dependent coupling strength $g(t)$ for $g_{0}t''=0.4$ and $g_{0}t''=0.1$, respectively. Other parameters are $J/g_{0}=5$, $g_{0}t'=0.1$, and $N=4$.}\label{fig4}
\end{figure}

We first consider the case of $N=4$ where the atom exhibits enhanced decay due to the constructive interference between the two coupling paths~\cite{LonghiGA,DLsyn}. Figure~\ref{fig4} depicts the dynamic evolutions of the atomic population in this case with different values of quench-off duration $t''$ and shows the evolution without coupling quench for comparison. In the absence of coupling quench, the atom exhibits a short-time parabolic decay before the subsequent exponential behavior (see the top inset) due to the strong memory effect of the structured bath. Based on this short-time behavior, one can find that the atomic decay slows down gradually and even tends to be inhibited as the quench-off duration $t''$ increases. 

Physically, this is because the memory of the lattice (i.e., the feedback from the lattice to the atom) tends to vanish for long enough quench off and the decay of the atom restarts with the short-time parabolic behavior whenever the couplings are turned on again. That is to say, a coupling quench with large enough quench-off duration mimics an ideal observation which results in a ``collapse'' of the state~\cite{OhZeno}. From this perspective, the turn-on duration $t'$ corresponds to the time interval between the observations, while the quench-off duration $t''$ serves as the duration of each observation. 

\begin{figure}
\includegraphics[width=8.5 cm]{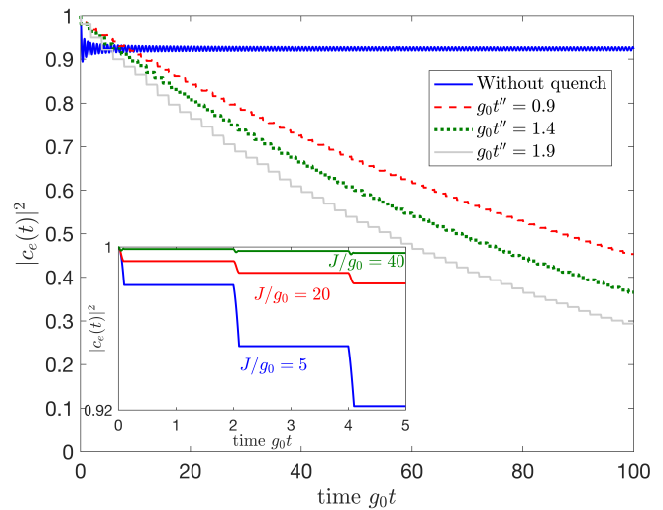}
\caption{Dynamic evolutions of atomic population $|c_{e}(t)|^{2}$ with initial state $|\psi(0)\rangle=\sigma_{+}|V\rangle$ in the case of a discrete waveguide. The inset depicts the evolutions of $|c_{e}(t)|^{2}$ for $g_{0}t''=1.9$ and different values of $J$. Other parameters are $J/g_{0}=5$, $g_{0}t'=0.1$, and $N=2$.}\label{fig5}
\end{figure}

Finally, we demonstrate in Fig.~\ref{fig5} that how periodic coupling quenches can be used to induce a long-lived giant atom to decay, simulating the QAZE of an open quantum system~\cite{AZEnature,FacchiPRL2001,FacchiJPA}. As shown in Fig.~\ref{fig5}, in the absence of coupling quench and for $N=2$, the population of the giant atom is in some sense undamped (with a persistent oscillation) after the retarded feedback term arising from the giant-atom structure takes into effect. To simulate the QAZE with the present model, we consider periodic coupling quenches with long enough quench-off durations (i.e., large enough $t''$) to avoid the memory effect of the lattice sites especially those between the two coupling points (for $N=2$, the energy can be partially confined between the two coupling points such that the feedback coming from the lattice cannot be ignored if the quench-off duration is short~\cite{DLsyn}). Therefore in Fig.~\ref{fig5}, we consider the cases of $g_{0}t''=\{0.9,\,1.4,\,1.9\}$, which demonstrate that the atomic population decays periodically in an exponential-like manner and falls to zero eventually. This is because the retarded feedback (and hence the giant-atom effect) never kicks in if $t'$ is smaller than $\tau$ and $t''$ is large enough. In view of this, the QAZE here should diminish gradually as the time delay $\tau$ approaches zero. This can be seen from the inset of Fig.~\ref{fig5}, where the atomic decay is almost suppressed if the model enters the Markovian regime with small enough $\tau$ (see, e.g., the green line with $\tau=d/v_{g}=N/2J=0.025/g_{0}$~\cite{LonghiGA,Longhiparadox}). Moreover, the present scheme is quite different from that of the small-atom case~\cite{OhZeno}, where the atom is assumed to be off-resonant with the lattice and its decoherence is partially suppressed due to the formation of an atom-photon bound state. 

\section{Conclusions}

To summarize, we have investigated the decay dynamics of a two-level giant atom with non-Markovian retardation effect and time-dependent atom-waveguide coupling strengths. We have revealed that the dynamic evolutions of the atomic population can be remarkably modified by the time-varying couplings, depending on the specific modulation form of the coupling strengths as well as the time delay of the retardation effect. Different from the case without modulation, here the retarded feedback term at time $t$ depends on the atom-waveguide coupling strengths at the earlier time $t-\tau$, where $\tau$ is the traveling time of photons between the two coupling points. This, thus, provides an alternative way for dynamically controlling the decay dynamics of the atom without changing the frequency of the emitted photons. In particular, by changing the coupling strengths abruptly, it is possible to realize a stationary population revival with which the atomic population grows to a higher value and stays there permanently. Moreover, we have extended our model to a discrete version, in which the atom can exhibit obvious short-time parabolic decay due to the strong memory effect of the structured bath, and have simulated the QZE and QAZE with the help of periodic coupling quenches. Different from the small-atom case, the present model exhibits the QAZE even if the transition frequency of the atom lies within the energy band of the lattice. These results not only have potential applications for controlling the decoherence effects in quantum networks, but also provide an alternative platform for studying non-Markovian retardation effects, Zeno physics, and quench dynamics in open quantum systems.

\section*{Acknowledgments} 

L.D. thanks Francesco Ciccarello, Quansheng Zhang, and Wu Wang for helpful discussions. This work is supported by the National Natural Science Foundation of China (under Grants No. 12074030, No. 11774024, and No. U1930402) and the Science Foundation of the Education Department of Jilin Province (under Grant No. JJKH20211279KJ).   

\appendix

\section{Derivation of Eq.~(\ref{eq6})}\label{appa}

Considering that the dispersion relation of the waveguide can be approximately linearized as $\omega_{k}\approx\omega_{0}+\nu=\omega_{0}+(k-k_{0})v_{g}$ around the transition frequency $\omega_{0}$ of the atom (due to the Weisskopf-Wigner approximation), Eq.~(\ref{eq5}) becomes
\begin{equation}
\begin{split}
\dot{c}_{e}(t)&=-\int_{0}^{t}dt'\int_{-\infty}^{+\infty}\Big\{2g_{1}(t)g_{1}(t')+2g_{2}(t)g_{2}(t')\\
&\quad\,+[g_{1}(t)g_{2}(t')+g_{1}(t')g_{2}(t)][e^{i(k_{0}+\nu/v_{g})d}\\
&\quad\,+e^{-i(k_{0}+\nu/v_{g})d}]c_{e}(t')e^{-i\nu(t-t')}\Big\}\frac{d\nu}{v_{g}}\\
&=-\int_{0}^{t}dt'\int_{-\infty}^{+\infty}\Big\{2g_{1}(t)g_{1}(t')+2g_{2}(t)g_{2}(t')\\
&\quad\,+[g_{1}(t)g_{2}(t')+g_{1}(t')g_{2}(t)][e^{ik_{0}d}e^{-i\nu(t-t'-d/v_{g})}\\
&\quad\,+e^{-ik_{0}d}e^{-i\nu(t-t'+d/v_{g})}]c_{e}(t')\Big\}\frac{d\nu}{v_{g}}.
\end{split}
\label{eqa1}
\end{equation}
According to the definition $\int d\omega\textrm{exp}(i\omega t)=2\pi\delta(t)$ of the $\delta$ function and its sifting property $\int dtf(t)\delta(t-t')=f(t')$, Eq.~(\ref{eqa1}) can be further simplified to
\begin{equation}
\begin{split}
\dot{c}_{e}(t)&=\frac{-2\pi}{v_{g}}\int_{0}^{t}dt'\Big\{[2g_{1}(t)g_{1}(t')+2g_{2}(t)g_{2}(t')]\\
&\quad\,\times\delta(t-t')+[g_{1}(t)g_{2}(t')+g_{1}(t')g_{2}(t)]\\
&\quad\,\times[e^{i\phi}\delta(t-t'-\tau)+e^{-i\phi}\delta(t-t'+\tau)]c_{e}(t')\Big\}\\
&=-\frac{2\pi}{v_{g}}[g_{1}(t)^{2}+g_{2}(t)^{2}]c_{e}(t)-\frac{2\pi}{v_{g}}e^{i\phi}\\
&\quad\,\times[g_{1}(t)g_{2}(t-\tau)+g_{1}(t-\tau)g_{2}(t)]\\
&\quad\,\times c_{e}(t-\tau)\Theta(t-\tau),
\end{split}
\label{eqa2}
\end{equation} 
where $\phi=k_{0}d$ and $\tau=d/v_{g}$ as defined in the main text. In the last step of Eq.~(\ref{eqa2}), we have discarded the time-advanced term containing $\delta(t-t'+\tau)$ due to its zero contribution to the integral.

\section{More details of the population revival and reduction}\label{appb}

For the steplike modulations shown in Fig.~\ref{fig3}, the dynamic equation of the atomic population for $t\in[t',t'+\tau)$ can be given by Eq.~(\ref{eq12}). When $\phi=(2m+1)\pi$, as shown in Fig.~\figpanel{fig3}{b}, the atomic population has reached a steady value ($\dot{c}_{e}=0$) before the sudden coupling change occurs, such that the instantaneous dynamic equation at $t=t'$ can be written as  
\begin{equation}
\begin{split}
\dot{c}_{e}(t=t')&=-\frac{4\pi g_{0}^{2}}{v_{g}}\frac{\Delta_{g}}{g_{0}}\Big(1+\frac{\Delta_{g}}{g_{0}}\Big)c_{e}(t)\\
&=-\Gamma_{0}F_{s}c_{e}(t).
\end{split}
\label{eqb1}
\end{equation} 
Note that $c_{e}(t'-\tau)=c_{e}(t')$ has been assumed here, which always holds as long as $t'>2\tau$. Clearly, one can expect an abrupt population revival (reduction) at $t=t'$, if the value of $F_{s}$ in the last step of Eq.~(\ref{eqb1}) is negative (positive). Then the atom becomes undamped again when the retarded feedback balances the transient process at $t=t'-\tau$, as shown in Eq.~(\ref{eq13}). Although Eq.~(\ref{eqb1}) fails to describe the case of $\phi\neq(2m+1)\pi$, one can still understand from Eq.~(\ref{eq12}) that the modified evolutions arise from the altered interference effect between the instantaneous and retarded terms at $t=t'$.

\end{document}